 \documentclass[doublespacing]{elsart}

 \usepackage{epsfig}

\usepackage{amssymb}

\begin{document}

\begin{frontmatter}



\title{Scaler Mode Technique for the ARGO-YBJ Detector}


\author{\centerline {{\it The Argo-YBJ Collaboration:}}}
\newline
\author[ROMA2,INFN_RM2]{G. Aielli},
\author[ROMA3,INFN_RM3]{C. Bacci},
\author[SALERNO,INFN_NA]{F. Barone},
\author[NAPOLI,INFN_NA]{B. Bartoli},
\author[LECCE,INFN_LE]{P. Bernardini},
\author[IHEP] {X.J. Bi},
\author[LECCE,INFN_LE]{C. Bleve \thanksref{cor2}},
\author[INFN_RM3]{P. Branchini}
\author[INFN_RM3]{A. Budano},
% ...
\corauth[cor1]{ corresponding author }

\address[ROMA2]{Dipartimento di Fisica dell'Universit\`a ``Tor Vergata''
                 di Roma, via della
                 Ricerca Scientifica 1, 00133 Roma, Italy}
\address[INFN_RM2]{Istituto Nazionale di Fisica Nucleare, Sezione di Tor 
        Vergata, via della Ricerca Scientifica 1, 00133 Roma, Italy}
\address[ROMA3]{Dipartimento di Fisica dell'Universit\`a ``Roma Tre'' di 
                 Roma, via della Vasca Navale 84,
                 00146 Roma, Italy}
\address[INFN_RM3]{Istituto Nazionale di Fisica Nucleare, Sezione di
                   Roma3, via della Vasca Navale 84, 00146 Roma, Italy}
\address[SALERNO]{Dipartimento di Scienze Farmaceutiche dell'Universit\`a
                 di Salerno, via Ponte Don Melillo, 
                 84084 Fisciano (SA), Italy}
\address[INFN_NA]{Istituto Nazionale di Fisica Nucleare, Sezione di
                 Napoli, Complesso Universitario di Monte
                 Sant'Angelo, via Cinthia, 80126 Napoli, Italy}
\address[NAPOLI]{Dipartimento di Fisica dell'Universit\`a di Napoli, 
                 Complesso Universitario di Monte
                 Sant'Angelo, via Cinthia, 80126 Napoli, Italy}
\address[LECCE]{Dipartimento di Fisica dell'Universit\`a di Lecce, 
                 via Arnesano, 73100 Lecce, Italy}
\address[INFN_LE]{Istituto Nazionale di Fisica Nucleare, Sezione di
        Lecce, via Arnesano, 73100 Lecce, Italy}
\address[IHEP]{Key Laboratory of Particle Astrophyics, Institute of High
                  Energy Physics, Chinese Academy of Science,
                  P.O. Box 918, 100049 Beijing, P.R. China}
\address[INFN_TO]{Istituto Nazionale di Fisica Nucleare,
        Sezione di Torino, via P.Giuria 1 - 10125 Torino, Italy}
% ...
\address[INFN_PV]{Istituto Nazionale di Fisica Nucleare Sezione di Pavia, 
        via Bassi 6, 27100 Pavia, Italy}
% ...
\address[INFN_CT]{Istituto Nazionale
        di Fisica Nucleare, Sezione di Catania, Viale A. Doria 6 -
        95125 Catania, Italy}
% ...

\begin{abstract}

The ARGO-YBJ experiment has been designed to study the Extensive Air
Showers with an energy threshold lower than that of the existing 
arrays by exploiting the high altitude location
(4300 m a.s.l. in Tibet, P.R. China) and the full 
ground plane coverage.
The lower energy limit of the detector (E $\sim$ 1 GeV) 
is reached by the scaler mode technique, i.e. recording the counting rate 
at fixed time intervals.
At these energies, transient 
signals due to local (e.g. Forbush Decreases) and cosmological
(e.g. Gamma Ray Bursts) phenomena are expected as a significant variation 
of the counting rate compared to the background.
In this paper the performance of the ARGO-YBJ detector operating
in scaler mode is described and discussed.

\end{abstract}

\end{frontmatter}


\section{Introduction}

The study of gamma ray sources in the 1 - 100 GeV energy range
has so far been only partially covered by observations by the
satellite-based EGRET, with a maximum detectable energy of 30 GeV.
On the other hand,
present ground-based detectors such as Cherenkov Telescopes are
still trying to reduce their threshold to energies well lower
than 100 GeV.
\\
In this energy range,
Gamma Ray Bursts (GRBs) are certainly the most intriguing transient phenomena.
GRBs have been deeply studied in the keV-MeV energy range, but only
EGRET, in the last decade, provided  some information in the GeV range.
Three bursts have been detected at energies $>$1 GeV,
with one photon of E=18 GeV \cite{catelli,hurley}.
The study of the high energy emission of GRBs could
provide extremely useful constraints to the emission models 
and the parameters of the surrounding medium.
\\
The interest for transient emission in the GeV region
is not only restricted to extragalactic
phenomena such as GRBs but also includes events closer to us - inside
our Solar System - such as
gamma rays from Solar Flares or high energy solar protons producing
short time increases of the detector counting rate at the
ground (Ground Level Enhancements).
\\
The study of these transient phenomena can be successfully
performed by ground-based experiments such as Extensive Air Shower (EAS)
detectors working in ``single particle mode''\cite{vernetto}. 
EAS arrays usually detect air showers generated by cosmic rays of
energy E $\geq$ (1 $-$ 100) TeV; the arrival direction of the
primary particle is measured by the time delay of the
shower front in different counters, and the primary energy is
evaluated by the number of secondary particles detected.
\\
EAS arrays can work in the energy region E $<$1 TeV
operating in single particle mode, i.e. counting all the particles
hitting the individual detectors, independently of whether they
belong to a large shower or they are the lone survivors of small
showers. Because of the cosmic ray spectrum steepness, most of the
events detected with this technique are in fact due to solitary
particles from small showers generated by 1 $-$ 100 GeV cosmic
rays.
\\
Working in single particle mode, an air shower array could in
principle detect a transient emission in the interval 1 $-$ 100 GeV (both by
gamma rays or charged primaries) if the secondary particles
generated by the primaries give a statistically significant excess
of events over the background due to cosmic rays.
\\
The power of this technique is in its extreme
simplicity: it is sufficient to count all the particles hitting
the detectors during fixed time intervals (i.e. every second or
more frequently, depending on the desired time resolution) and to
study the counting rate behaviour versus time, searching for
significant increases. The observation of an excess in coincidence
with a GRB detected by a satellite would be an unambiguous signature of
the nature of the signal.
\\
The single particle technique does not allow one to measure 
energy and direction of the primary gamma rays, because
the number of detected particles (often only one per shower) is too small
to reconstruct the shower parameters. However, it can allow to
study the temporal behaviour of the high energy emission, and,
with some assumptions on the spectral slope (possibly supported by
satellite measurements at lower energies), can give an evaluation
of the total power emitted.
\\
The background rate depends on the altitude and, to a
lesser extent, on the geomagnetic latitude; it is
not strictly constant and several effects both environmental and
instrumental are responsible for
variations on different time scales with amplitudes up to a few
per cent.
For this technique, an accurate knowledge of the detector and its
sensitivity to both environmental and instrumental parameters is of crucial
importance in order to correctly evaluate the statistical significance of
the detected signals. 
\\
Obviously the sensitivity of this technique increases with the
detection area and the energy threshold decreases at higher
observation levels, so the most suitable detectors have a large area and
work at very high altitude.
\\
The ARGO-YBJ experiment, located at the YangBaJing High Altitude
Cosmic Ray Laboratory (Tibet, P.R. China, 4300 m a.s.l.) with a
detection area of $\sim$6700 $m^2$, successfully exploits the full
coverage approach at very high altitude with the aim of studying
the cosmic radiation with a low energy threshold.
\\
In order to cover many physics items in different energy ranges,
two main operation
modes have been designed: (1) shower mode,
with an energy threshold of a few hundreds of GeV, and
(2) scaler mode, with a threshold of a few GeV, this latter differing
from the single particle mode for different particle
multiplicities (i.e. not only the single particles) are put in
coincidence and counted on different scaler channels.
\\
The large field of view ($\gtrsim$2 $sr$, limited only by atmospheric 
absorption) and the
high duty cycle (close to 100$\%$) typical of EAS detectors, make
ARGO-YBJ suitable to search for unknown sources and unpredictable events
such as GRBs or other transient phenomena.
The site location
(latitude 30$^{\circ}$ 06${'}$ 38${''}$ N, longitude 90$^{\circ}$
31${'}$ 50${''}$ E) allows sky monitoring in the declination 
band $-10^{\circ}<\delta<70^{\circ}$.
\\
In this paper the performance of ARGO-YBJ working in scaler mode
will be described, with some examples of application of this technique
in solar physics and in the search for GeV emission from GRBs.

\section{The Detector}

The ARGO-YBJ experiment is made by a single layer of Resistive Plate Chambers 
(RPCs) \cite{nim_rpc} housed in a large building (100$\times$110 $m^2$).
The detector has a modular structure: the basic module is a cluster
(5.7$\times$7.6 $m^2$), made of 12 RPCs (2.850$\times$1.225 $m^2$
each).
130 of these clusters are organized in a full coverage carpet of 5600 $m^2$
with active area $\sim$93\%; this central detector is surrounded by 23 
additional clusters with a coverage of $\sim$40\% (``guard ring'') to improve 
the core location reconstruction.
The installation of the central carpet was completed in June 2006.
\\
Each RPC is read via 80 strips (6.75$\times$61.80 $cm^2$), logically 
organized in 10 pads of 55.6$\times$61.8 $cm^2$ that are 
individually recorded and that represent the space-time pixels of the 
detector.
The RPC carpet is connected to two different data acquisition systems, 
working independently, and
corresponding to the two operation modes, shower and scaler.
In shower mode, for each event the location and timing of every detected
particle is recorded, allowing the lateral distribution and arrival
direction reconstruction. In scaler mode the total counts are 
measured every 0.5 s:
for each cluster, the signal coming from the 120 pads is added up and put
in coincidence in a narrow time window (150 ns), giving the counting rates
of $\ge$1, $\ge$2, $\ge$3, $\ge$4 pads, that are read by
four independent scaler channels.
The corresponding measured counting rates are $\sim$ 40 kHz, $\sim$ 2 kHz,
$\sim$ 300 Hz and $\sim$ 120 Hz.
\\
In order to correctly handle the data, it is very important to evaluate the
response to particles hitting the detector.
For scaler mode operations the most important parameters are the detector 
efficiency and the strip cross-talk, i.e., the probability of having adjacent 
pads fired by a single particle, giving false coincident counts.
These effects strongly depend on the adopted gas
mixture and power supply (in our case Argon 15\%, 
Isobutane 10\% and Tetrafluoroethane 75\% at 7200V).
In these conditions a detector efficiency $\epsilon$ = 95\% has been
measured together with a strip occupancy, i.e., the mean number of strips
fired by a single particle, of 1.2 \cite{nim_rpc}.
A very simple analytical calculation has been performed to evaluate the 
contribution of single particles to higher multiplicities: for the given 
strip occupancy, there is a probability of 10\% for each strip to generate 
double counts.
A ``cross-talk region'', approximated for sake of simplicity to a step 
function, is defined as a width corresponding to 10\% of the strip span.
Taking into account the measured detection efficiency, the possibility of
having cross-talk on both short and long strip sides, and the front-end 
digital logic - giving a maximum number of counts for each pad equal to 1 
independently of the number of particles hitting simultaneously the pad -
we obtain the following probabilities of measuring 0, 1 or 2 pads from a 
single particle hitting the RPC: p(0) = 5\%; p(1) = 93\%; p(2) = 2\%
(our analytical approximation does not 
give multiplicities greater than 2).
This result has been checked experimentally on-site with one chamber, 
finding: p(0) = 5.8\%; p(1) = 92.0\%; p(2) = 2.1\%;
p($\ge$3) = 0.04\% \cite{icrc_rpc}.
The good agreement confirms the validity of our approach.
\\
Finally, a substantially different behaviour is expected by the single
particle channel ($\ge1$) with respect to the higher multiplicity channels,
because of the contributions given by spurious
dark counts (related to front-end electronics
and chamber noise) and natural radioactivity.
Since from both we expect mostly single counts, 
these contributions should influence only the $\ge$1 scaler 
due to the very low rate of random coincidences.
An on-site measurement to disentangle all these contributions to the single 
particle rate has been scheduled.

\section{Detector Stability}

Several effects are known to modify the background
counting rate due to cosmic rays.
First of all, variations in the atmospheric pressure
affect the secondary particle flux: an increase (decrease) of the
ground level atmospheric pressure results in a reduction
(enhancement) of the background rate, because of the larger
(smaller) absorption of the electromagnetic component.
By monitoring the pressure at the detector position it is possible 
to correct the data for this effect. 
Local phenomena connected to the temperature and/or to the wind \cite{wind}
at the level of the early shower development have also been measured, 
the effect being a change of the probability of secondary pions
to decay with respect to their interaction, due to the local density variation.
These effects are expected to be
negligible compared to the above mentioned barometric effect, and very 
difficult to measure since the outdoor temperature, that is usually
monitored, is weakly correlated to the high atmosphere 
($\sim 10$ $km$) temperature.
Moreover, the experiment hall temperature  modifies the
detector performance, changing the gas density inside the RPCs
and the bakelite electrodes resistivity.
\\
The cosmic ray background variations due to these parameters 
are not negligible (up to a few per cent of the counting rate in some cases),
nevertheless for signals shorter than a few minutes (such as a high
energy emission from a GRB) they can be neglected due to 
the very different time scales of the two phenomena.
Much more troublesome are possible 
instrumental effects, such as electric noise, that could 
simulate a GRB signal producing spurious increases in the 
background rate. Working in single particle mode requires very 
stable detectors, and a very careful and continuous monitoring of 
the experimental conditions. By comparing the counting rates of the 
scaler channels and requiring simultaneous and consistent 
variations in all of these, it is possible to identify 
and reject most of the noise events due to instrumental effects. 
In the following subsection the study of the detector behaviour
over short time periods is discussed.
\\
On the other hand, when the study involves long duration signals
(as for solar activity or sidereal anisotropies) a correction 
to the counting rates is required in order to feature the underlying phenomena.
The effects of our corrections for the atmospheric pressure and the hall 
temperature and their efficiency in shrinking the counting rate distributions
towards a Poissonian distribution are discussed in section 3.2.

\subsection{Detector stability over short time periods}

\begin{figure}[ht] 
\begin{center}
\includegraphics[width=32.pc]{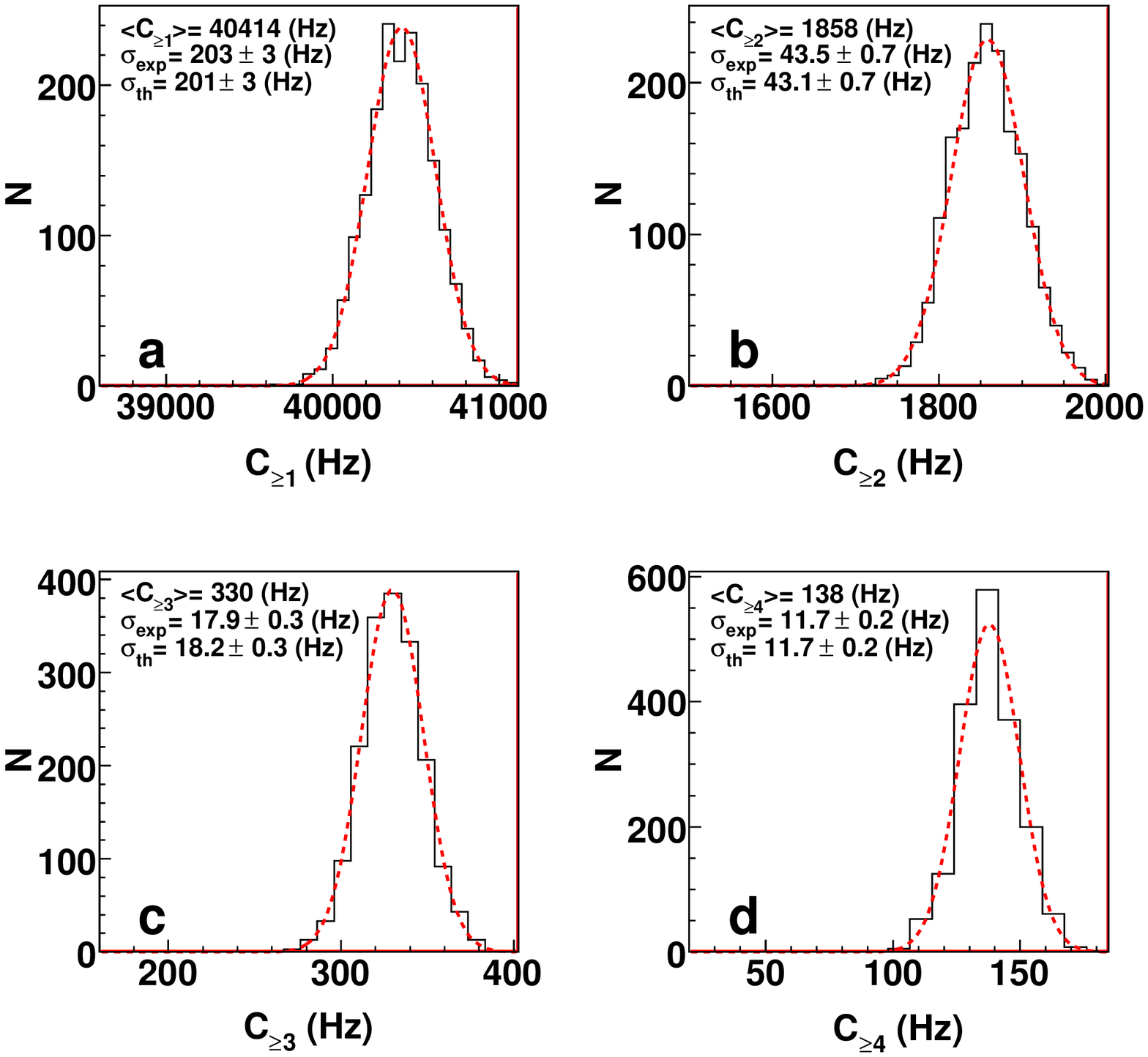} 
\end{center}
\caption{Experimental distributions of the counting rates  
and their Gaussian fits for a typical cluster: (a) $C_{\ge1}$, 
(b) $C_{\ge2}$, (c) $C_{\ge3}$ and (d) $C_{\ge4}$ for
30 minutes data accumulation. The standard deviation of the Gaussian fit
($\sigma_{exp}$) is compared with the square root of the mean 
of the experimental distribution ($\sigma_{th}$) 
to check the compatibility with the Poisson distribution.}
\label{shorttime1} 
\end{figure}

\begin{figure}[ht] 
\begin{center}
\includegraphics[width=16.pc]{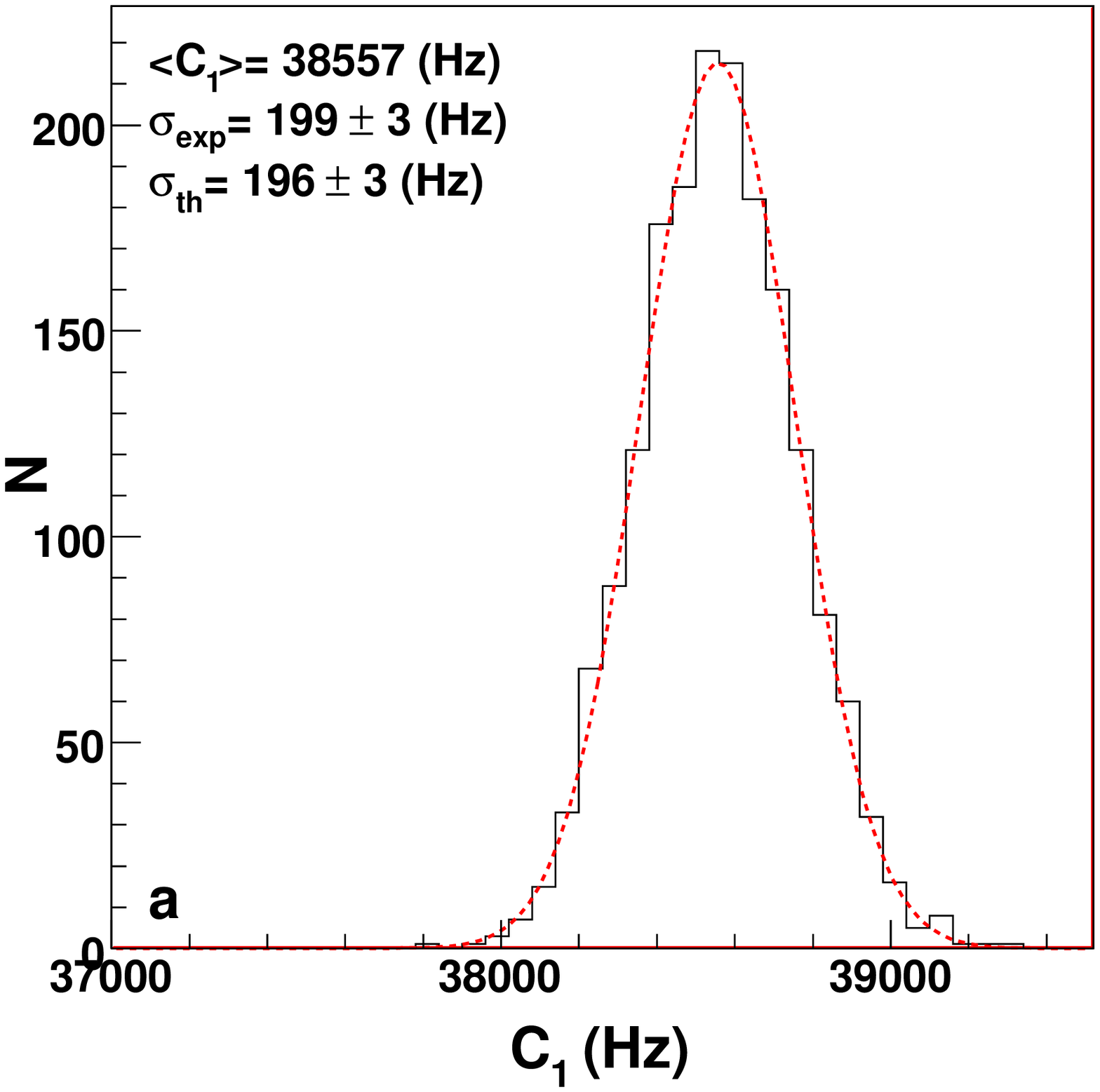} 
\includegraphics[width=16.pc]{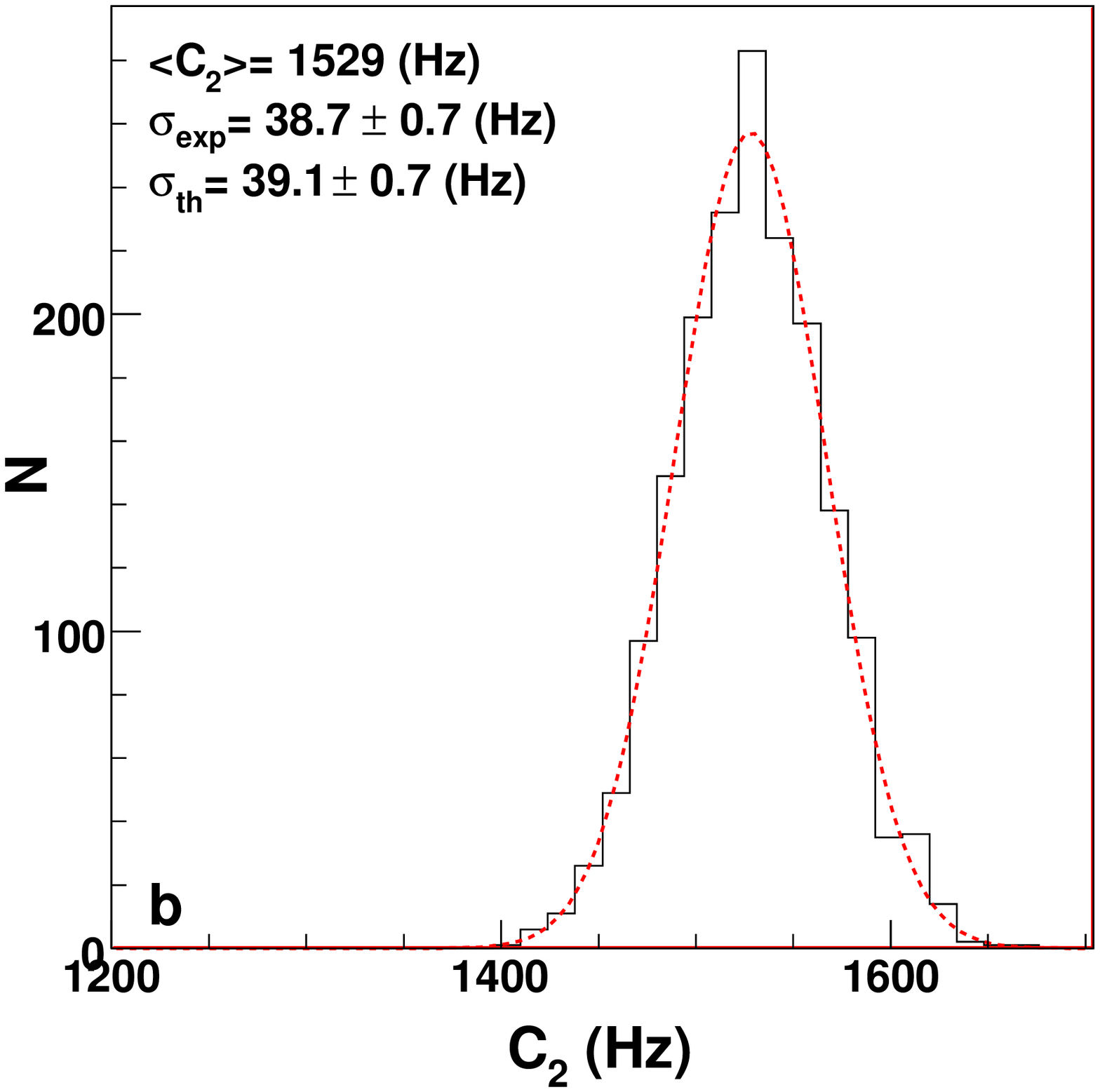} 
\includegraphics[width=16.pc]{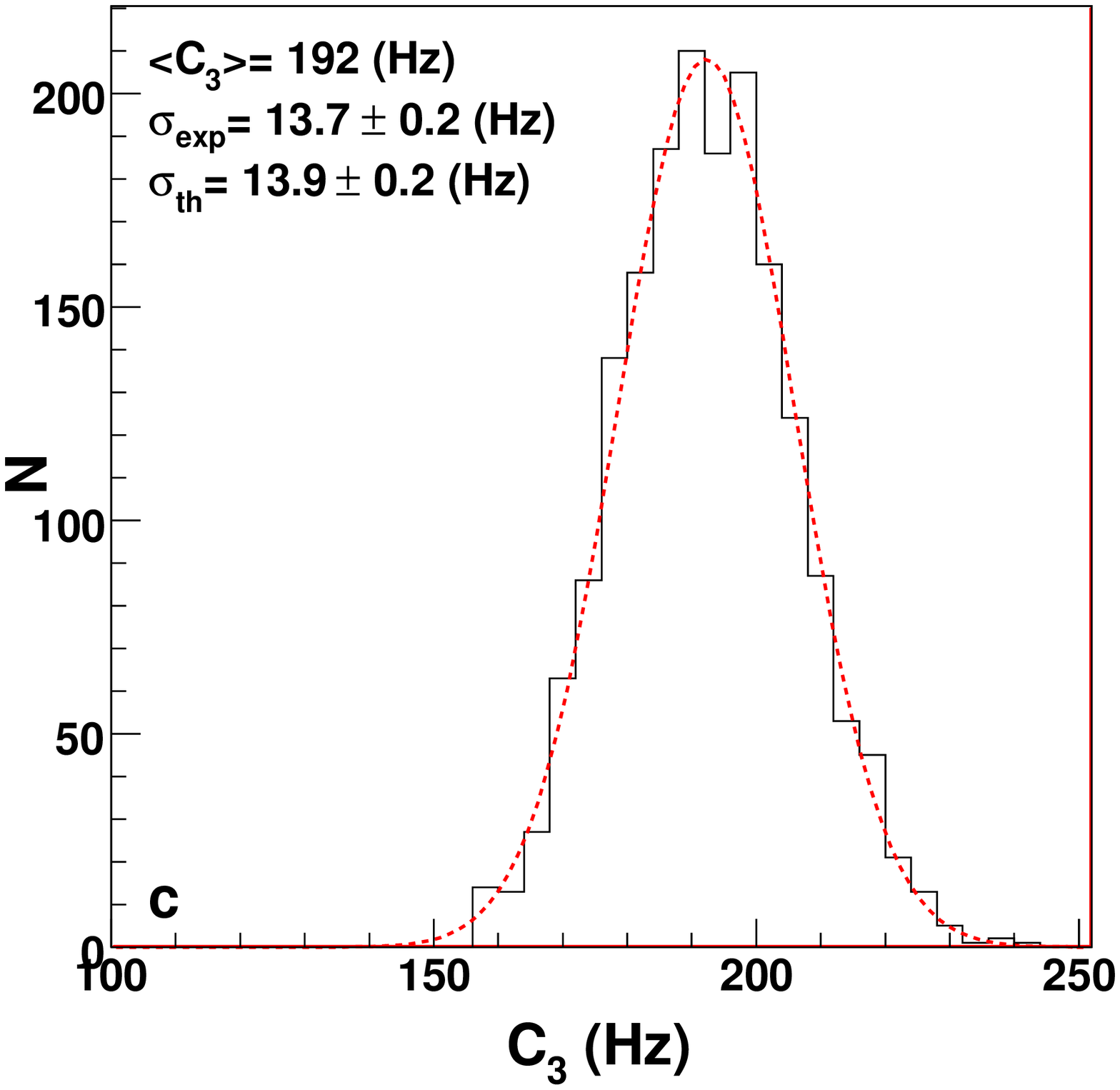} 
\end{center}
\caption{Experimental distributions of the counting rates
and their Gaussian fits for a typical cluster: (a) $C_1$, 
(b) $C_2$ and (c) $C_3$ for 30 minutes  
data accumulation. The standard deviation of the Gaussian fit
($\sigma_{exp}$) is compared with the square root of the mean of
the experimental distribution ($\sigma_{th}$) 
to check the compatibility with the Poisson distribution.}
\label{shorttime2} 
\end{figure}

We define as ``short'' a time period with a duration less than half an hour.
Figs. \ref{shorttime1} and \ref{shorttime2} show the counting rates for 
multiplicities $\ge$1, $\ge$2, $\ge$3, $\ge$4
and 1, 2, 3 for a typical cluster during 30 minutes of data 
acquisition.
The counting rates $C_i$ are obtained from the measured counting rates
$C_{\ge i}$ using the relation:

\begin{equation}
C_i = C_{\ge i} - C_{\ge i+1} \; \; \; (i=1,2,3)
\end{equation}

The distributions for all $C_{\ge i}$ and $C_i$ are Poissonian, meaning that 
the environmental parameter variations have negligible effects over such a 
time scale.
A useful variable is obtained by adding up counting rates of different 
multiplicities 
for the same cluster or different clusters for the same counting rate 
multiplicity.
In this case, non-Poissonian behaviours appear due to 
the correlation between different channels.
For the total number of particles hitting a cluster we have:

\begin{equation}
C_{tot} = \sum_{i=1}^{4} i C_i \simeq C_{1}+2C_{2}+3C_{3}+4C_{\ge4} =
C_{\ge1} + C_{\ge2} + C_{\ge3} + C_{\ge4}
\label{tot_counts}
\end{equation}

\begin{figure}[ht] 
\begin{center}
\includegraphics[width=32.pc]{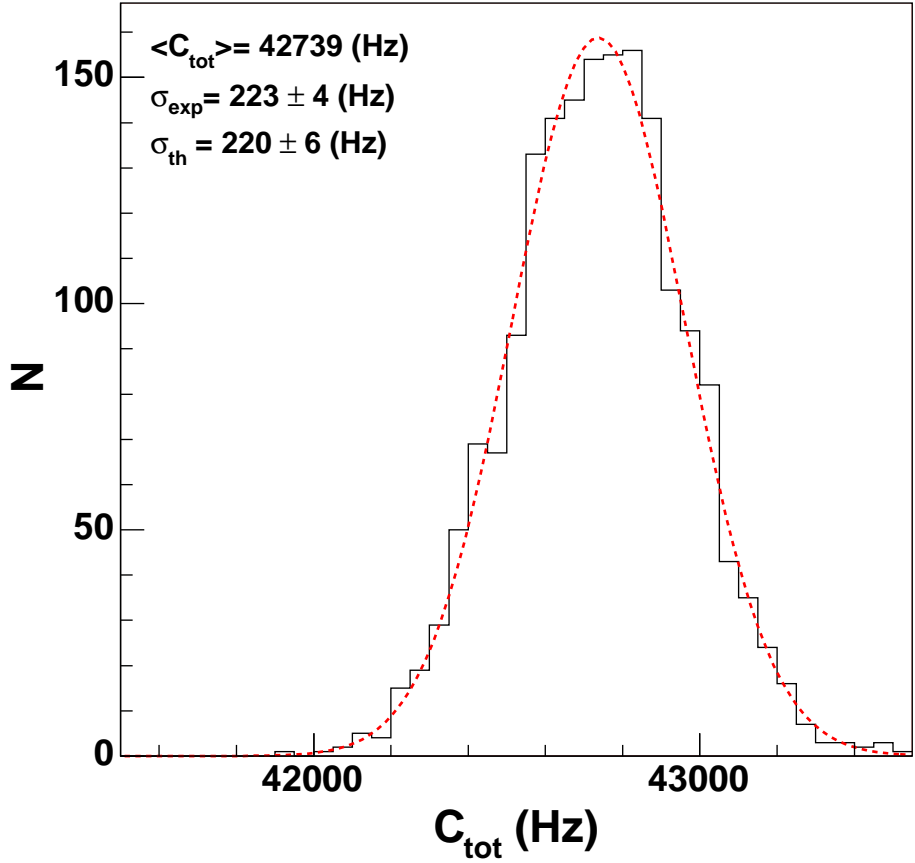} 
\end{center}
\caption{Experimental distribution of the total number of
particles hitting a typical cluster and its Gaussian fit for 30
minutes data accumulation. The standard deviation of the Gaussian fit
($\sigma_{exp}$) is compared with the expected value derived from the
single multiplicities counting rates (see text, eq. (\ref{wsum})).}
\label{scalsum} 
\end{figure}

\begin{figure}[ht] 
\begin{center}
\includegraphics[width=32.pc]{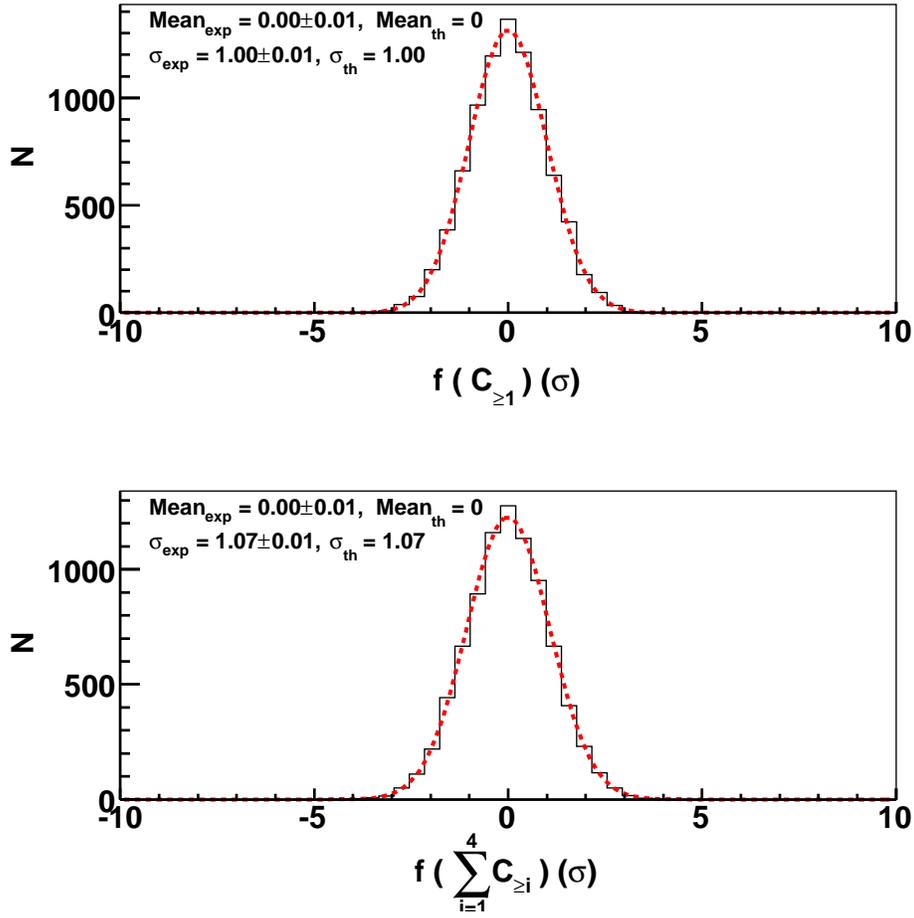} 
\end{center}
\caption{Experimental distribution of the normalized excesses 
of signal over 
background of a typical cluster (see text, eq. (\ref{ffunction})) 
compared with a Gaussian fit.
Top: $C_{\ge1}$ channel; bottom: sum of the four channels.}
\label{fdistrib} 
\end{figure}

Applying the error propagation to eq. (\ref{tot_counts}) we obtain:

\begin{equation}
\sigma^2_{C_{tot}} = \langle C_{1}\rangle+4\langle C_{2}\rangle+
                     9\langle C_{3}\rangle+16\langle C_{\ge4}\rangle 
\label{wsum}
\end{equation}

that is larger than a Poisson distribution.
Fig. \ref{scalsum} shows the experimental distribution of 
the counting rate $C_{tot}$ 
for the same cluster and time interval shown in 
figures \ref{shorttime1} and \ref{shorttime2}.
As expected, the distribution is Gaussian, wider than Poissonian and 
with a width $\sigma_{exp}$=(223$\pm$4) Hz, compatible with the value of 
(220$\pm$6) Hz obtained by applying eq. (\ref{wsum}).
\\
More difficult to estimate is the correlation between different clusters 
due to the probability that some of the counts are generated by the same 
events giving the same multiplicity in different clusters.
These events mean the counts in different clusters are not independent,
widening the distribution from Poissonian with
the sum of counts from more clusters.
An analytical calculation gives the following result: if $M_i$ is the counting
rate of cluster $i$ for a given multiplicity ($\ge$1,
$\ge$2, $\ge$3, $\ge$4 or 1, 2, 3) 
and $n$ is the total number of clusters, we have:

\begin{eqnarray}
M_{tot} & = & \sum_{i=1}^{n} M_i \nonumber \\
\sigma^2_{M_{tot}} & = 
& \langle M_{tot}\rangle \left(1+\sum_{i=2}^{n} (i^2 -i) p_i\right) 
\label{csum}
\end{eqnarray}
 
where $p_i$ is the probability that the multiplicity is
satisfied for the same event on (and only on) $i$ clusters.
As an example, if we consider three clusters and multiplicity $\ge$1,
we can have two or three clusters hit by at least one particle coming from
the same shower.
In this case, $p_3 = \frac{N_{123}}{M_{tot}}$ and 
$p_2 = \frac{N_{12}+N_{13}+N_{23}}{M_{tot}}$, where $N_{123}$ is the number
of counts generated by events hitting all the three clusters
(with at least one particle) and $N_{ij}$ by events hitting one couple of
clusters.
$N_{ij}$ is not included in $N_{123}$: if three clusters are hit by the
same event this contributes only to $N_{123}$.
\\
The width of the $M_{tot}$ distribution is larger than the  
Poissonian value
$\sigma^2_{M_{tot}} = \langle M_{tot}\rangle$; the $p_i$ values depend on the 
given multiplicity, 
the EAS lateral distribution function and the cluster layout.
Since the correlation between different clusters is topologically-dependent,
for the same number of clusters and multiplicity we
expect a wider distribution for close clusters 
compared to more distant ones.
This has been checked experimentally. As the sensitivity of the scaler 
mode technique is inversely proportional to background fluctuations, 
a continuous carpet is at a disadvantage with respect to scattered
detectors with the same total surface area.
\\
The behaviour is the same for the distribution of the ratio between 
signal and background fluctuations (instead of the count rates), as 
in the GRB search analysis.
Considering a number $s$ of counts in a time interval $\Delta$t
corresponding to the GRB duration and a 
background $b$ of counts  
averaged on a time interval 10 times $\Delta$t before and after the signal, 
we define the normalized fluctuation function:

\begin{equation}
f = (s-b)/\sigma, \; \; \; \; \; 
\sigma = \sqrt{b+b/20}
\label{ffunction}
\end{equation}

which represents the significance of the excess compared to background 
fluctuations.
\\
Fig. \ref{fdistrib} shows the $f$ distribution for the same cluster 
of Figs. \ref{shorttime1}-\ref{scalsum} and for 
$\Delta$t = 10 $s$, obtained using one day of data, for  
multiplicity $\ge1$ and for the sum of the four multiplicities from
$\ge1$ to $\ge4$.
Notice that for the single channel the Poissonian behaviour of the $f$ 
distribution is preserved even if the total distribution is obtained with 24 
hours of data, provided that the total time duration of signal plus 
background is less than 30 minutes.
Since the value of $f$ is calculated independently for each 
signal and background time intervals,
the long term variations can be neglected and the single channel response
is always ``locally'' Poissonian.
On the contrary, the $f$ distribution of the sum $\sum_{i=1}^{4} C_{\ge}i$ 
follows equation (\ref{wsum}), and the 
width of 1.07 can be easily obtained by using the experimental counting rates.
The fact that the distribution of $f$ is not affected by long term variations 
is also very useful to select good and bad channels, the latter showing
very clearly anomalous non-Poissonian fluctuations.

\subsection{Detector stability over long time periods and correlation with
environmental parameters}

\begin{figure}[ht] 
\begin{center}
\includegraphics[width=32.pc]{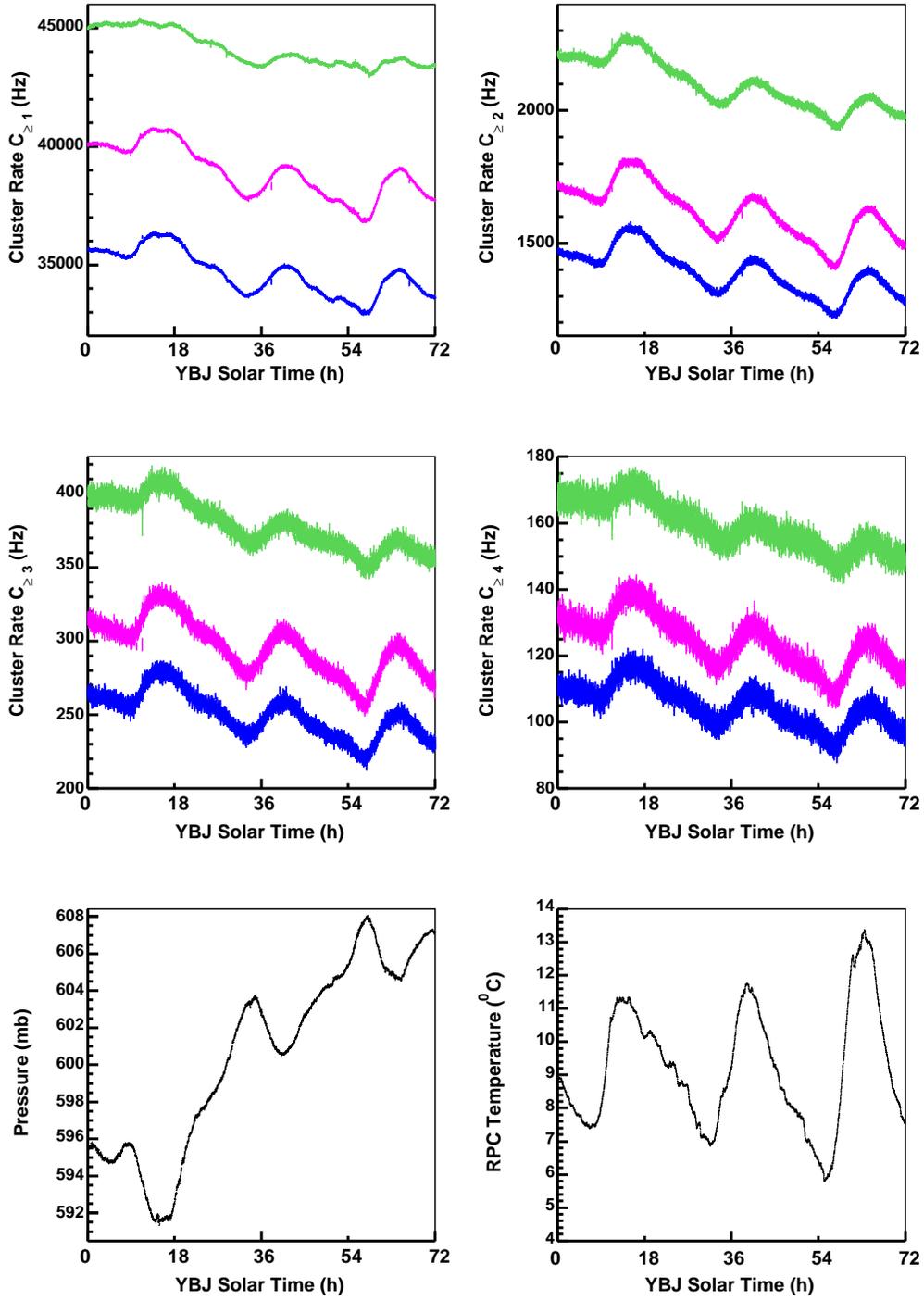} 
\end{center}
\caption{Counting rates as a function of time for 3 clusters 
for multiplicities from
$\ge1$ to $\ge4$ and 72 hours of data accumulation, together with pressure (P) 
and RPC temperature (T) behaviour for the same time interval (lower plots).}
\label{countings} 
\end{figure}

\begin{figure}[ht] 
\begin{center}
\includegraphics[width=32.pc]{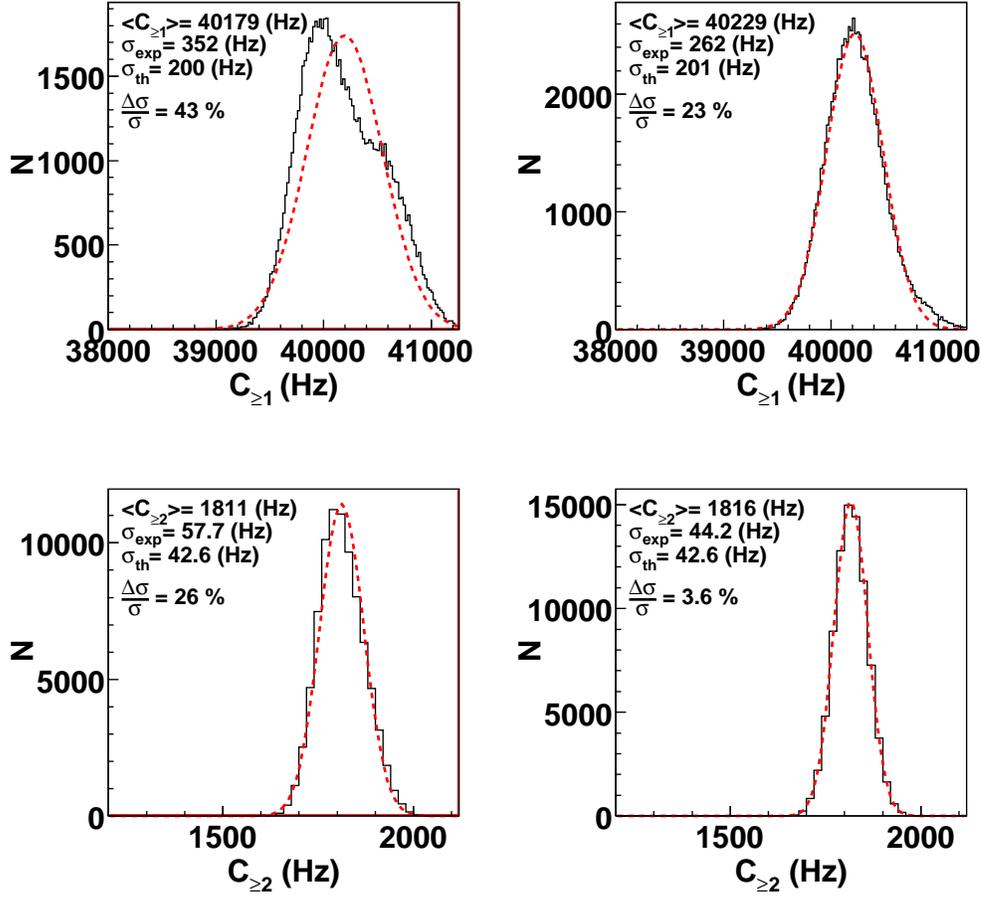}  
\end{center}
\caption{Experimental distributions of counting rates 
for $C_{\ge1}$ and $C_{\ge2}$ 
of a typical cluster before (left) and after (right) corrections for 
pressure and RPC temperature in a period of 24 hours.
The standard deviation of the Gaussian fit
($\sigma_{exp}$) is compared with the square root of the mean of 
the experimental distribution ($\sigma_{th}$) and the 
resulting ``non-Poissonian'' 
behaviour $\frac{\Delta\sigma}{\sigma}=\frac{\sigma_{exp}-\sigma_{th}}
{\sigma_{exp}}$ is reported.}
\label{corr_pt} 
\end{figure}

As already stated at the beginning of this section, several effects are 
expected to modulate the counting rate measured by our RPC detector, 
affecting both the shower development and instrumental response.
These effects must be corrected before any analysis of long term phenomena, 
i.e. with duration greater than $\sim$ 30 minutes.
\\
Changes in the atmospheric pressure (P) and the RPC temperature (T)
are expected to dominate the count rate 
variations.
The correlation with the atmospheric pressure has been widely studied, and
a typical coefficient of $\approx$ 0.7\%/mbar is found for a very large 
range of primary energies \cite{corr_p}.
At our atmospheric depth (606 $g\; cm^{-2}$), the vertical flux of soft ($e$)
and hard ($\mu$) components are almost equivalent \cite{hillas},
reducing the dependency of the pressure correction coefficient
on the atmospheric depth (i.e. zenith angle) due to the muonic component.
The connection between the counting rates and the local temperature is mainly 
due to the anti-correlation of the bakelite electrode plates resistivity
with the temperature \cite{Arna,Abbre}.
In addition, a variation in the detector temperature modifies the gas density, 
causing a change in the effective gap width 
(the gas volume being constant): the
final result is a direct correlation between the counting rate and the 
detector temperature, that 
we assume is the temperature of the gas just out of the RPC.
\\
Fig.\ref{countings} shows the counting rates for 3 clusters for 
multiplicities from $\ge1$ to $\ge4$ as a function of local solar time for 
72 hours of data accumulation, together with the P and T behaviour.
Besides the atmospheric pressure and the detector temperature, the solar 
diurnal anisotropy \cite{Jacklyn} is responsible for the counting rate 
modulation, while the differences in efficiency, chamber dark current
and electronics noise give the offset displacements.
\\
In order to correct for these effects we performed a fit with a two-dimensional
function (as a first approximation linear in P, T) given by the following 
expression:

\begin{equation}  
C(P,T) \simeq C_0 [1-\mu(P-P_0)] [1+\beta(T-T_0)] 
\label{atmpress}
\end{equation} 

where $P_0$ and $T_0$ are the reference values of pressure and temperature 
for counting rate $C_0$, and $\mu$ and $\beta$ are respectively the 
barometric and thermal coefficients.
For the counting channel $\ge1$ we found a barometric coefficient 
$\mu$ = 0.3-0.5\%/mbar and
a thermal coefficient $\beta$ = 0.2-0.4\%/$^\circ$C, depending on the cluster 
considered  and the specific experimental conditions.
For the other 
scalers we obtain $\mu$ = 0.9-1.2\%/mbar and 
$\beta$ = 0.2-0.4\%/$^\circ$C \cite{pTargo}.
\\
Fig. \ref{corr_pt} shows the counting rate distribution for $C_{\ge1}$ and 
$C_{\ge2}$ before and after corrections for the previous parameters in a 
period of 24 hours.
It can be seen that for the counting channel $\ge1$ the corrected data becomes
Gaussian but not yet Poissonian, with an experimental standard deviation
23\% greater than the expected value.
In contrast, the distribution of $C_{\ge2}$ is Gaussian for 
the raw data and becomes Poissonian after corrections.
The fact that the two raw distributions for the same day and the 
$\mu$, $\beta$ coefficients are completely different for $C_{\ge1}$ and 
$C_{\ge2}$ may be explained assuming that part of the $C_{\ge1}$ counts are 
not due to cosmic rays.
\\
The statistical behaviour of the counting rates after corrections 
for atmospheric pressure and RPC temperature has been 
checked for longer periods showing an agreement better than 
5\% up to 1 month for 
$C_{\ge2}$ or higher scalers, while for $C_{\ge1}$ the distribution becomes 
asymmetrical and much larger ($\sim$36\%) than Poissonian already after 48 
hours.

\subsubsection{The Forbush Decrease of January 2005}

\begin{figure}[ht] 
\begin{center}
\includegraphics[width=32.pc]{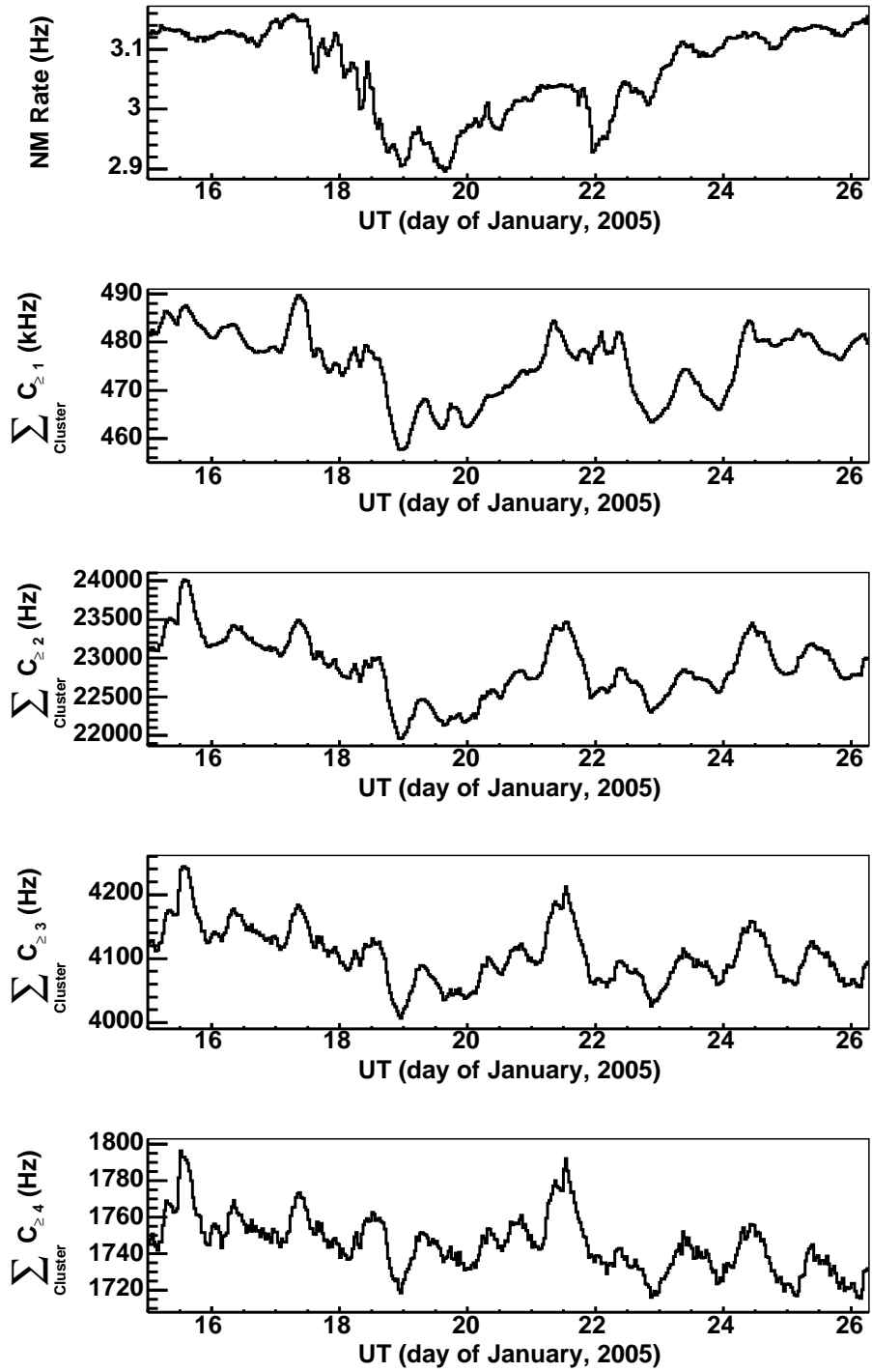} 
\end{center}
\caption{Plot of the YBJ neutron monitor (NM) data compared with the 
four counting rate distributions (added on all the operating 12 clusters and
corrected only for the 
atmospheric pressure) during the Forbush Decrease of January 2005
(Neutron Monitor data: courtesy of the Yangbajing Neutron Monitor 
Collaboration).}
\label{FD} 
\end{figure}

An example of detection of a long-term phenomenon, linked to solar 
activity, is given by the Forbush Decrease (FD) observed in coincidence 
with the YBJ neutron monitor in January 2005.
A FD is the fall in the cosmic ray intensity at the Earth 
one or more days after an Earth directed Coronal Mass Ejection (CME) lift-off,
and is  
caused by the shielding effect of the magnetic field 
and shock associated with the high speed 
plasma cloud emitted from the Sun at the time of the eruption.
Large solar flares can be 
associated with Ground Level Enhancements (GLEs) and, depending on the
flare location, may also produce a CME that causes a FD; both of these 
phenomena are normally 
detected at the ground by neutron monitors, but a large area detector working 
in scaler mode is also suitable.
\\
The FD mentioned above was recorded by the worldwide network of neutron
monitors as a significant decrease in the counting rate after a sudden 
storm commencement.
The IGY neutron monitor at Jungfraujoch (Switzerland, vertical effective
cut-off rigidity $R_c$=4.5 GV) observed a decrease with onset 
around noon UT on January 17, 2005, 
and with a maximum amplitude of about -15\% \cite{IGY_Suisse}.
On January 20, during the recovery phase of the FD at Earth, an
X7.1 solar flare created a very intense burst of energetic paticles and a 
GLE was observed, with onset time between 6:48 and 6:57 UT and
maximum amplitude $>$ 5000\% at the South Pole neutron monitor station.
The GLE spectrum was very soft, with a spectral index approaching 
asymptotically $\alpha \sim$ -5 \cite{Bieber},
producing quite small responses at even moderate geomagnetic cut-off
rigidity locations like Jungfraujoch, where the maximum amplitude was 
11.4\%.
FDs similarly produce an energy dependent drop in intensity deeper at
low cut-off rigidity 
locations, but since their spectra are much harder than in GLEs
they should be observable by the ARGO-YBJ array, characterized by 
$R_c$=14 GV \cite{Storini}.
Fig. \ref{FD} shows the YBJ neutron monitor data compared 
with the four counting rate distributions corrected only 
for the atmospheric pressure since the RPC temperature measurement 
was not yet implemented.
Only 12 clusters were in operation at that time.
Even if the scaler mode counting rate is more sensitive to 
environmental parameters, a correlation can be clearly seen, with an amplitude 
of -7\% in the YBJ neutron monitor data and about -5\% and -4\% for $C_{\ge1}$ 
and $C_{\ge2}$ data.
No decrease is observed for $C_{\ge3}$ and $C_{\ge4}$, and no GLE is seen in 
any scaler channel and in the YBJ neutron monitor data.
Another interesting feature of FDs is the recovery phase.
While the magnitude of FDs depends on the cut-off rigidity (as already shown),
the recovery time is expected to be independent of the galactic cosmic ray
energy and thus similar for all detectors.
Despite of this, as pointed out in \cite{Jamsen},
the recovery time in some events strongly depends on the median energy
of the detected particles, 
with a faster recovery for more energetic cosmic rays.
Moreover, these higher energy cosmic rays show an "over-recovery", i.e.
a recovery of the cosmic ray intensity to a level higher than that before 
the FD.
The study of these features is very important to characterize and model 
the propagation of the interplanetary shocks; due to its high cut-off 
rigidity, ARGO-YBJ can contribute to the study of acceleration
processes, keeping in mind that the energy upper limit for incoming 
solar particles in the Earth environment has not yet been established.
Due to its huge area, this study can be done with high sensitivity and 
small integration time.
For example, a distinct enhancement of the diurnal variation in the
recovery phase after the FD can be seen mainly around 24-25 January.
It is most obvious in the $C_{\ge2}$ data, but it is also evident
in the $C_{\ge3}$ and $C_{\ge4}$ data where the FD is not observed.
This phenomenon is known but not extensively studied, hence our
data can give valuable information on the energy
dependency of the FD recovery.
A more detailed analysis of these data and their correlation is in progress.

\section{Effective Area and Expected Counting Rates}

The effective area is usually obtained by simulating a large number of
showers $N_{sim}$ with the core sampled on a wide area $A_{sim}$,
much wider than typical shower dimensions, and counting the number of
showers $N_{trig}$ matching the trigger condition:

\begin{equation}  
A_{eff}(E) = A_{sim} \cdot N_{trig}/N_{sim}
\label{effarea}
\end{equation} 

\begin{figure}[ht]
\begin{minipage}[ht]{.47\linewidth}
  \begin{center}
\includegraphics[width=18.pc]{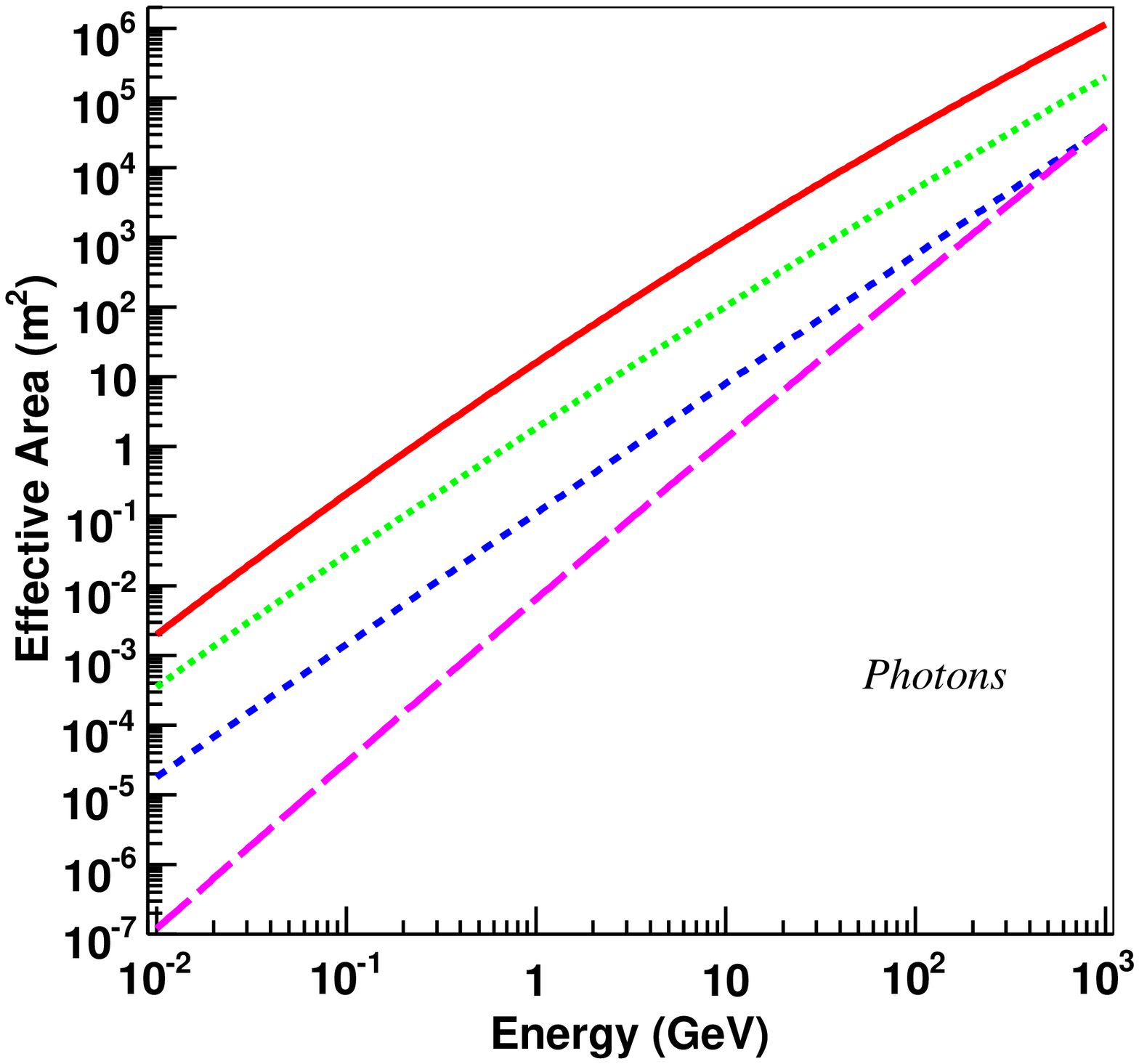} 
  \end{center}
\end{minipage}\hfill
\begin{minipage}[ht]{.47\linewidth}
  \begin{center}
\includegraphics[width=18.pc]{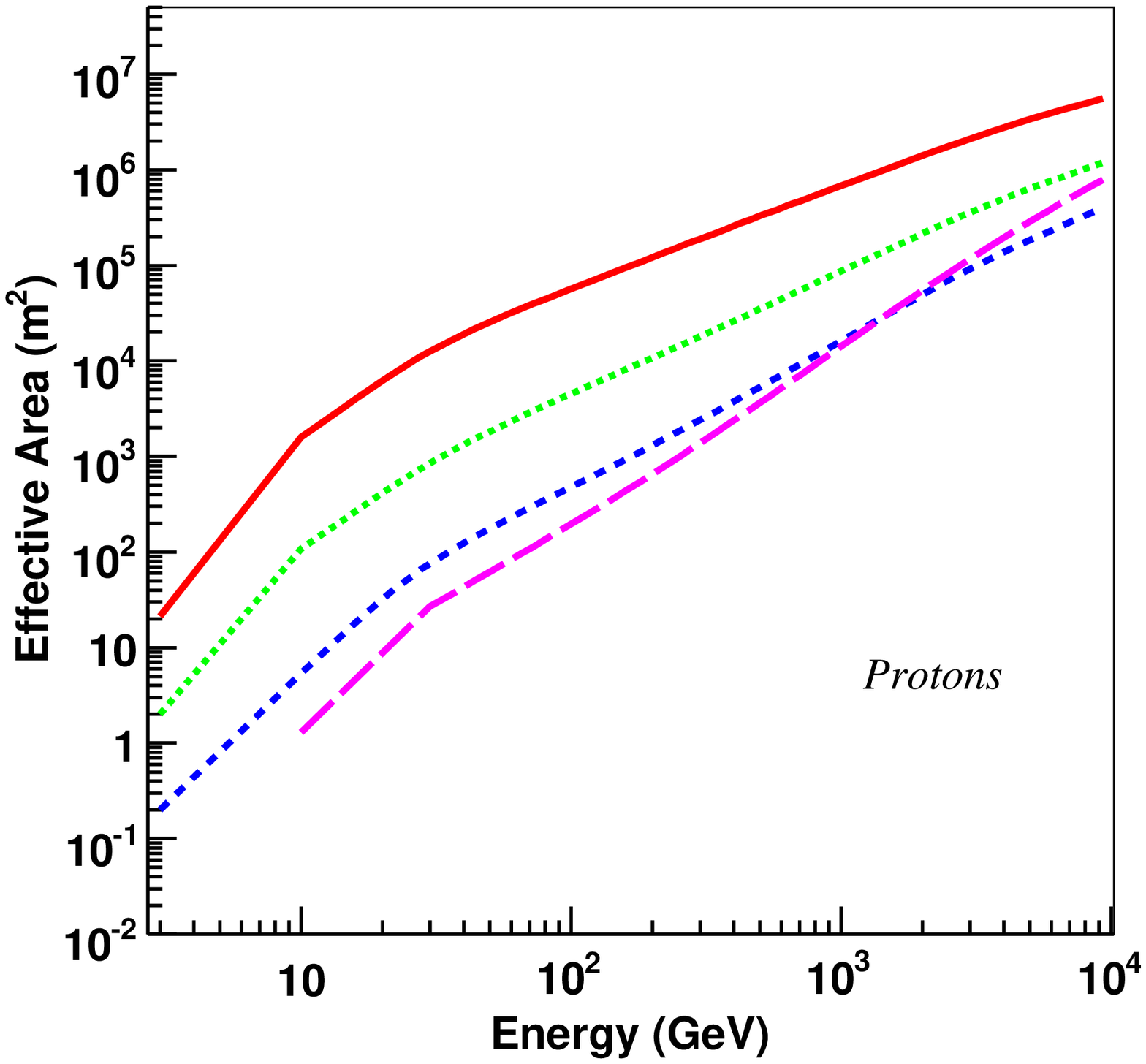} 
 \end{center}
\end{minipage}\hfill
    \caption{Effective areas for photon and proton-induced showers 
(left and right plot, respectively) with 
zenith angle $\theta$ = 20$^{\circ}$ 
in the four multiplicity channels 
(from top to bottom, 1, 2, 3 and $\ge4$).}
   \label{aeffi}
\end{figure}

\begin{figure}[ht]
\begin{minipage}[ht]{.47\linewidth}
  \begin{center}
\includegraphics[width=18.pc]{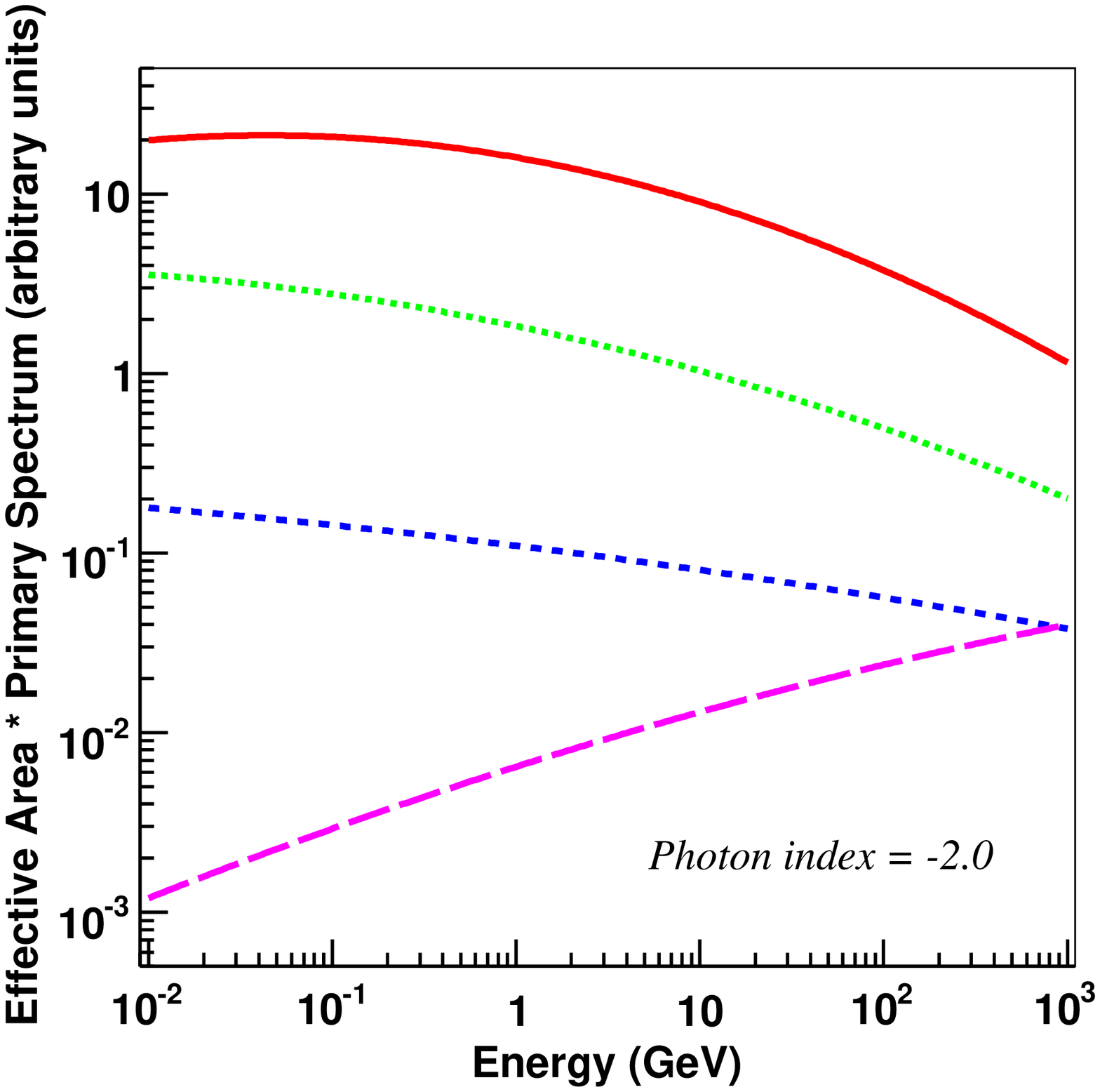} 
  \end{center}
\end{minipage}\hfill
\begin{minipage}[ht]{.47\linewidth}
  \begin{center}
\includegraphics[width=18.pc]{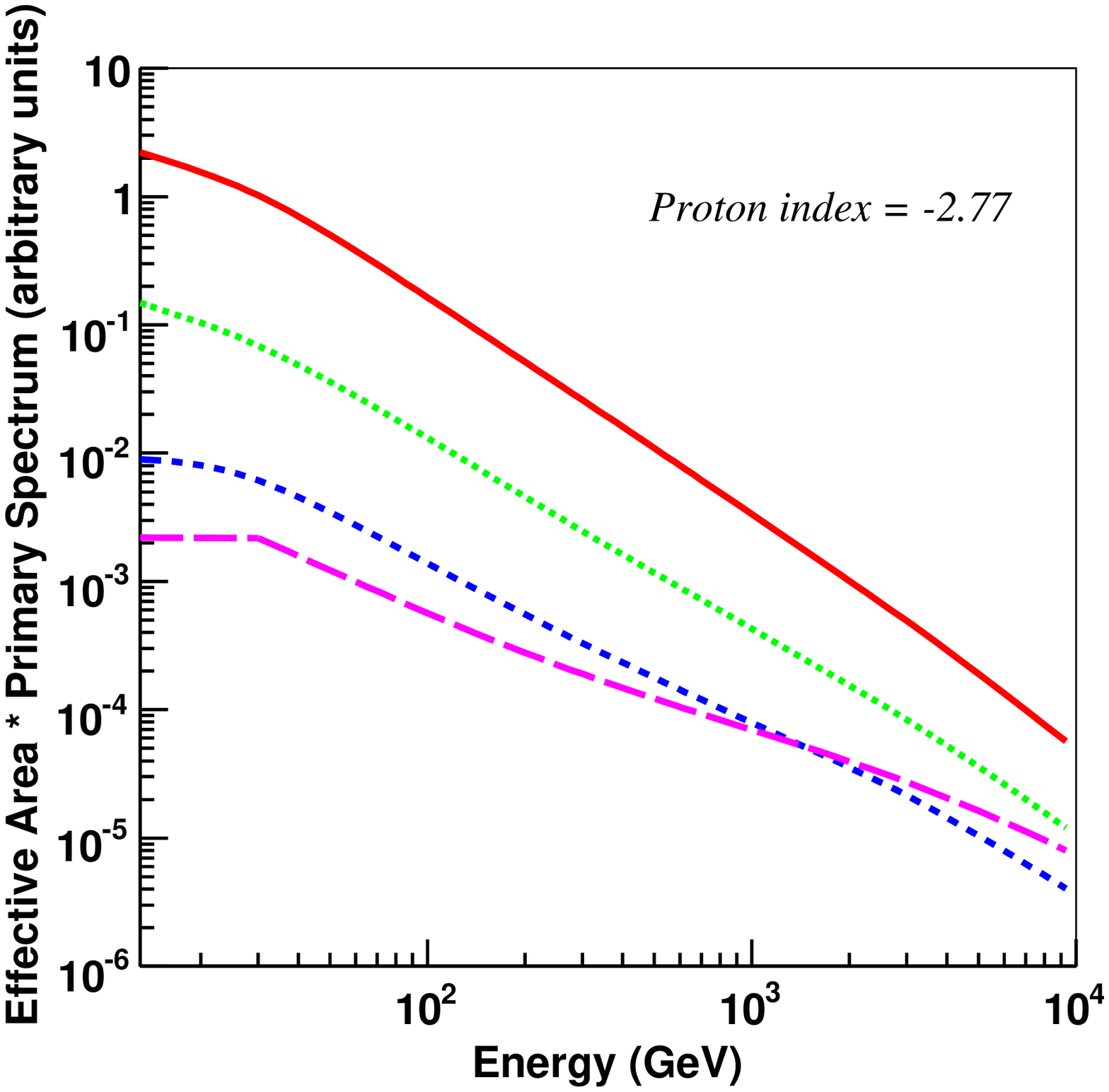} 
 \end{center}
\end{minipage}\hfill
    \caption{Convolutions of the effective areas of Fig. \ref{aeffi}
with a differential power law spectrum of index $\alpha$ = -2.0 for photons 
and $\alpha$ = -2.77 for protons, taking into 
account in this latter case the geomagnetic bending, in the four
multiplicity channels
(from top to bottom, 1, 2, 3 and $\ge4$).}
   \label{convol}
\end{figure}

For the ARGO-YBJ detector, the effective area has been estimated by means
of a detailed Monte Carlo (MC) simulation for photons, protons and helium
of fixed energies ranging from 10 MeV to 100 TeV 
and fixed zenith angle 
$\theta$ from 0 to 50 degrees (with 10 degrees steps). 
The CORSIKA/QGSJet code 6.204
\cite{corsika} has been used with a full electromagnetic component
development down to E$_{thr}$= 1 MeV for electrons and
photons and 50 MeV for muons and hadrons.
\\
Since the actual efficiency depends essentially on the shower
particle lateral distribution, a huge number of showers must be
simulated over an area large enough to completely contain the event.
\\ 
In order to save computing time the shower sampling has been performed 
by means of the {\em ``reciprocity technique''} \cite{battist}: the 
sampling area A$_S$ is uniformly filled with replicas of the same 
ARGO-YBJ carpet, one next to the other, for a total area
of $\sim 10^4 \times 10^4$ $m^2$. 
Following the reciprocity technique, we sample the shower axis only 
over the area covered by the carpet located at the center of the array, 
with the prescription of considering the response of all the detector
replicas. 
Then, event by event, we calculate the number of
clusters on the entire grid which contain at least 1, 2, 3, 4 fired pads.
In this calculation the cross-talk between adjacent pads has been taken
into account, as discussed in section 2.
The resulting effective areas for the full ARGO-YBJ detector 
as a function of the energy of primary photons and protons at zenith
angle $\theta$ = 20$^{\circ}$ are
shown in Fig. \ref{aeffi} in the four multiplicity channels
(1, 2, 3 and $\geq4$). 
\\
If the probability that more than one particle per shower
hits the detector is negligible, i.e. for sufficiently
small detector and/or small primary energy,
the effective area $A_{scaler}$ scales like the mean electron 
size after the shower maximum at the detection level \cite{vernetto}:
\begin{equation}\label{aeff-spt}
A_{scaler}\approx 
\epsilon\cdot \langle N_{Ch}\rangle\cdot A_d\cdot cos\>\theta.
\end{equation}
where $\epsilon$ and $A_d$ are the detector efficiency and area, and 
$\langle N_{Ch}\rangle$ is the mean number of charged secondary particles. 
>From Fig. \ref{aeffi} we see that, as expected, the dependence is described
by a power law index $\approx$ 1.6 for multiplicity 1, and this index 
increases with 
the multiplicity approaching the value $\approx$ 2.6, valid only in the 
shower trigger configuration \cite{abdo}.
The index lowers with increasing energy when the formula
(\ref{aeff-spt}) loses its validity because the area of the cluster 
is too large for the given shower particle density and thus more than 
one particle fires it.
The effective area for protons is usually lower than that for
photons of the same energy, increasing the sensitivity of the measurement,
except in the 10 -- 100 GeV energy range, where, due essentially to the
muonic component, the total number of charged particles in
proton-induced showers is larger than that in photon events.
Moreover, the geomagnetic cutoff at YBJ is $\geq$14 GV 
(depending on the arrival direction) \cite{Storini}.
\\
In order to get the counting rates, the effective area is folded with the
primary spectrum, expressed by a differential power
law with spectral index $\alpha$ = -2.0 for photons, 
$\alpha$ = -2.77 for protons and $\alpha$ = -2.64 for helium
\cite{barbara}.
We adopted $\alpha$ = -2.0, which is
the mean spectral index for GRBs as measured by EGRET \cite{dingus}.
The resulting convolutions for photons and protons are shown in 
Fig. \ref{convol}.
\\
The behaviour of these curves shows how each energy decade contributes to the
counting rate for a fixed zenith angle. 
One can note that using four scalers instead of a single one allows the 
investigation of different energy ranges, with the possibility of 
giving an indication on the source spectral index in case of positive 
detection.
\\
The resulting counting rates, integrating the different effective 
areas for angular bins up to a zenith angle of 60
degrees and both proton and helium contributions, 
are 24.4 kHz for $C_1$, 1.8 kHz
for $C_2$, 195 Hz for $C_3$ and 117 Hz for $C_{\ge4}$.
The uncertainties on these rates come mainly from those affecting the
primary fluxes which are estimated to be about 20\% (see for example
\cite{gaisser}).
A comparison with the measured rates, i.e., 38 kHz for $C_1$, 1.7 kHz
for $C_2$, 180 Hz for $C_3$ and 120 Hz for $C_{\ge4}$, shows 
a very good agreement between the predicted values and the experimental,
apart from the counting rate $C_1$ affected by spurious signals 
as explained in section 2.

\section{Upper Limits and Expected Sensitivity}

\begin{figure}[ht]
\begin{center}
\includegraphics[width=32.pc]{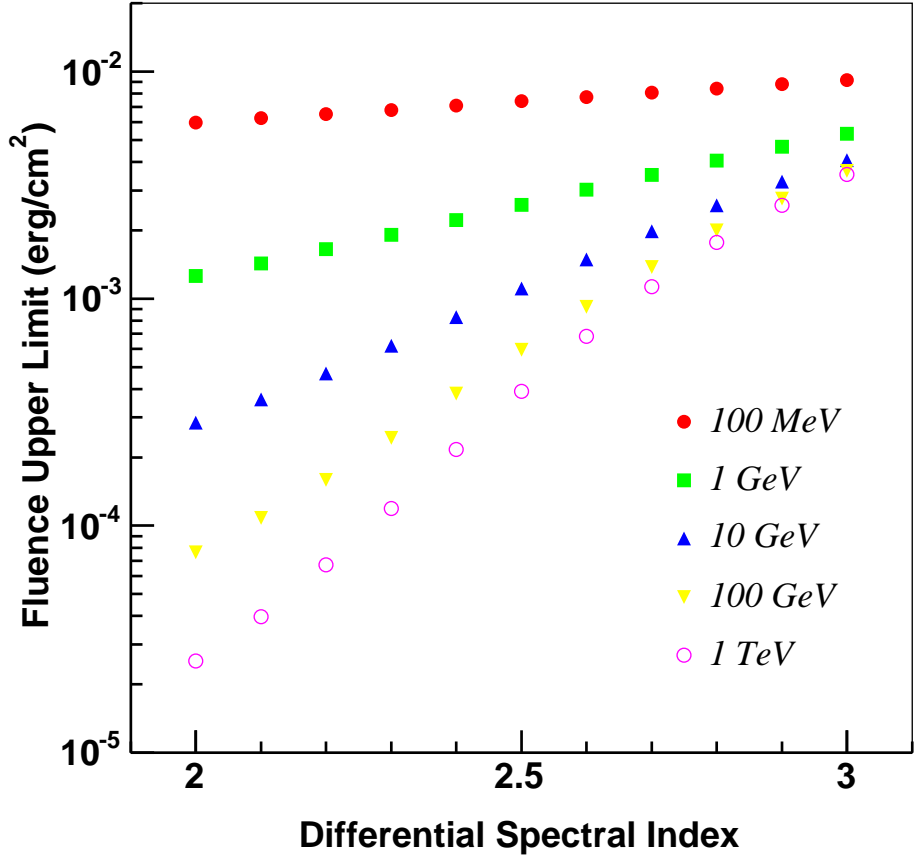}
\end{center}
\caption{Fluence upper limits in the energy range 
10 MeV - $E_{max}$ corresponding to a $4\sigma$ excess 
in the multiplicity channel 1, as a function of the GRB spectral index, 
for a zenith angle $\theta$=20$^{\circ}$, a time duration
$\Delta t=10\;s$ and a redshift $z$=0.
Each set of points refers to different $E_{max}$ as shown in the figure.}
\label{upperlim}
\end{figure}

The energy fluence $F$ in the energy range $[E_{min},E_{max}]$
corresponding to an
excess of $N$ events above the cosmic ray background is given by:

\begin{equation}
F(\alpha,E_{max}) = \frac{N\int_{E_{min}}^{E_{max}} E^{\alpha +1} dE}
{\int_{E_{min}}^{E_{max}}E^{\alpha} A_{eff} (E,\theta) dE}
\end{equation}

where $A_{eff} (E,\theta)$ is the effective area calculated in a given 
multiplicity channel at the fixed zenith angle $\theta$ and $\alpha$
is the photon spectral index (see previous section). 
\\
In case of no significant excess, typical fluence upper limits to 
GRB emissions 
are obtained using the measured counting rates in each of the four 
multiplicity channels and a model for the GRBs. 
\\
Assuming a power law spectrum up to $E_{max}$ ranging from 100 $MeV$
to 1 $TeV$, no $\gamma\gamma$ absorption by the Extragalactic 
Background Light, a zenith angle $\theta =
20^{\circ}$ and a duration $\Delta t=10\;s$, the upper limits corresponding 
to a $4\sigma$ excess in the multiplicity channel 1
are calculated as a function of the spectral index 
in the energy range 10 $MeV-E_{max}$.
The results are reported in Figure \ref{upperlim}. 
For different time durations, upper limits scale as $\sqrt{\Delta t(s)/10}$.
\\
In order to evaluate the real possibilities of a GRB detection, 
we compare the ARGO-YBJ sensitivity with the GRB fluxes measured by EGRET.
During its lifetime EGRET detected 16 events at energies $>$200 
MeV\cite{catelli}.
Above this energy,
all their spectra show a power law behaviour without any cutoff,
suggesting the possibility that a large part of GRBs could emit GeV or even
TeV gamma rays.
\\
In our simple model every EGRET GRB spectrum is extrapolated to 
higher energies, using the measured power law index, up to a maximum energy
$E_{max}$ representing a possible intrinsic cutoff at the source.
Then we consider an exponential cutoff to take into account the effects of the 
extragalactic absorption, obtaining the spectrum:
$dN/dE$ = $K$ $E^{\alpha}$ e$^{-\tau(E,z)}$. 
The intrinsic energy cutoff is assumed in the range
100 GeV $< E_{max} <$ 1 TeV and the GRB redshift in the range
0$\leq z \leq$3. The absorption factor is calculated by using the values of
optical depth $\tau(E,z)$ given in \cite{kneiske}.
\\
Our calculation shows that a fraction ranging from 10\% to 50\% of 
EGRET GRBs is above the ARGO-YBJ minimum detectable flux, 
depending on the cutoff energy $E_{max}$ and the redshift $z$ of the GRB.
Assuming $E_{max}$=1 TeV ($E_{max}$=100 GeV) and no absorption of gamma rays, 
ARGO-YBJ could detect 50\% (30\%) of EGRET GRBs.
If z=1 (z=3) this value reduces to $\sim$ 20\% (10\%), for both values of
$E_{max}$.
Considering that EGRET observed 16 GRBs during 7 years (1991-1997) and that
the EGRET field of view ($\sim$ 0.5 $sr$) is about four times smaller than
that of the present Swift satellite, $\sim$ 9 GRBs similar to those seen by 
EGRET are expected to occur in one year in the Swift field of view.
\\
Since about 1/10 of these bursts are, by chance, in the ARGO-YBJ field of view,
we expect $\sim$ 1 EGRET-like GRB detected by Swift 
to occur in our detector field of view in one year. 
A fraction of them (depending on $z$ and $E_{max}$)
will be effectively detectable.

\section{Conclusions}

The ARGO-YBJ detector operated in scaler mode has a high duty cycle
and optimal stability, 
fulfilling the fundamental requirements of the scaler mode technique.
\\
The main goal is the measurement of the high energy tail of GRB spectra, 
and the detection of few events during the experiment life time is expected 
if the spectrum of the most energetic GRBs extends at least up to 100 GeV.
In this case, with four measurement channels sensitive to different 
energies, valuable information on the high energy spectrum
slope and possible cutoff may be obtained.
\\
In any case, even if no signal is detected, the large field of view 
and the high
duty cycle, allowing for  continuous monitoring overhead,
make the ARGO-YBJ experiment one of the most sensitive detectors for the
study of the high energy spectrum of GRBs, with typical fluence upper
limits down to $10^{-5} erg/cm^2$ in the 1 - 100 GeV region, well
below the energy range explored by the present generation of Cherenkov
telescopes.
\\
On the other hand, due to the high rigidity cut-off, the large area of the
detector and its high mountain location allow the study of the solar
modulation of galactic cosmic rays and energetic relativistic particles
coming from the Sun with high sensitivity and fine time resolution.
This has been shown by the observation of the January 2005 FD with a 
preliminary discussion on the recovery phase.
\\
Data collected up to May 2008 have been analyzed 
according to the procedures outlined in this work and the results of
these analyses will be presented in later papers.

\section{Acknowledgements}

This paper is supported in part by the National Natural Science Foundation
of China (NSFC) under the grant No. 10120130794, the Chinese Ministry
of Science and Technology, the Key Laboratory of Particle Astrophysics, 
Institute of High Energy Physics (IHEP), Chinese Academy of Science (CAS) and 
the National Institute of Nuclear Physics of Italy (INFN).
\\
The authors thank M. Storini for useful discussion, M. Dattoli also 
thanks the National Institute of Astrophysics of Italy (INAF)
for partly supporting her activity in this work.




\newpage
\section{Captions list}
Fig. \ref{shorttime1} Experimental distributions of the counting rates  
and their Gaussian fits for a typical cluster: (a) $C_{\ge1}$, 
(b) $C_{\ge2}$, (c) $C_{\ge3}$ and (d) $C_{\ge4}$ for
30 minutes data accumulation. The standard deviation of the Gaussian fit
($\sigma_{exp}$) is compared with the square root of the mean 
of the experimental distribution ($\sigma_{th}$) 
to check the compatibility with the Poisson distribution.\\
Fig. \ref{shorttime2} Experimental distributions of the counting rates
and their Gaussian fits for a typical cluster: (a) $C_1$, 
(b) $C_2$ and (c) $C_3$ for 30 minutes  
data accumulation. The standard deviation of the Gaussian fit
($\sigma_{exp}$) is compared with the square root of the mean of
the experimental distribution ($\sigma_{th}$) 
to check the compatibility with the Poisson distribution.\\
Fig. \ref{scalsum} Experimental distribution of the total number of
particles hitting a typical cluster and its Gaussian fit for 30
minutes data accumulation. The standard deviation of the Gaussian fit
($\sigma_{exp}$) is compared with the expected value derived from the
single multiplicities counting rates (see text, eq. (\ref{wsum})).\\
Fig. \ref{fdistrib} Experimental distribution of the normalized excesses 
of signal over 
background of a typical cluster (see text, eq. (\ref{ffunction})) 
compared with a Gaussian fit.
Top: $C_{\ge1}$ channel; bottom: sum of the four channels.\\
Fig. \ref{countings} Counting rates as a function of time for 3 clusters 
for multiplicities from
$\ge1$ to $\ge4$ and 72 hours of data accumulation, together with pressure (P) 
and RPC temperature (T) behaviour for the same time interval (lower plots).\\
Fig. \ref{corr_pt} Experimental distributions of counting rates 
for $C_{\ge1}$ and $C_{\ge2}$ 
of a typical cluster before (left) and after (right) corrections for 
pressure and RPC temperature in a period of 24 hours.
The standard deviation of the Gaussian fit
($\sigma_{exp}$) is compared with the square root of the mean of 
the experimental distribution ($\sigma_{th}$) and the 
resulting ``non-Poissonian'' 
behaviour $\frac{\Delta\sigma}{\sigma}=\frac{\sigma_{exp}-\sigma_{th}}
{\sigma_{exp}}$ is reported.\\
Fig. \ref{FD} Plot of the YBJ neutron monitor (NM) data compared with the 
four counting rate distributions (added on all the operating 12 clusters and
corrected only for the 
atmospheric pressure) during the Forbush Decrease of January 2005
(Neutron Monitor data: courtesy of the Yangbajing Neutron Monitor 
Collaboration).\\ 
Fig. \ref{aeffi} Effective areas for photon and proton-induced showers 
(left and right plot, respectively) with 
zenith angle $\theta$ = 20$^{\circ}$ 
in the four multiplicity channels 
(from top to bottom, 1, 2, 3 and $\ge4$).\\
Fig. \ref{convol} Convolutions of the effective areas of Fig. \ref{aeffi}
with a differential power law spectrum of index $\alpha$ = -2.0 for photons 
and $\alpha$ = -2.77 for protons, taking into 
account in this latter case the geomagnetic bending, in the four
multiplicity channels
(from top to bottom, 1, 2, 3 and $\ge4$).\\
Fig. \ref{upperlim} Fluence upper limits in the energy range 
10 MeV - $E_{max}$ corresponding to a $4\sigma$ excess 
in the multiplicity channel 1, as a function of the GRB spectral index, 
for a zenith angle $\theta$=20$^{\circ}$, a time duration
$\Delta t=10\;s$ and a redshift $z$=0.
Each set of points refers to different $E_{max}$ as shown in the figure.\\

\clearpage
\section{Figures}

\clearpage
 
\clearpage
 
\clearpage

\clearpage

\clearpage

\clearpage

\clearpage

\clearpage

\clearpage

\end{document}